\documentclass[english,a4paper]{article}
\usepackage{babel,amssymb,graphicx,hyperref}

\advance\textheight \topskip \textwidth 35.5pc \oddsidemargin 20pt
\begin{document}

\author{V.~Strokov$^{1,2}$\footnote{{\bf e-mail}: strokov@asc.rssi.ru},
\\
$^{1}$ \small{\em Astrospace Center of the P.~N.~Lebedev Physical Institute of RAS} \\
\small{\em 117997, Moscow, ul. Profsoyuznaya, 84/32} \\
$^{2}$ \small{\em Moscow Institute of Physics and Technology,}\\
\small{\em Department of General and Applied Physics} \\
\small{\em 141701, Dolgoprudny, Institutsky per., 9} }
\title{On convergence to equilibrium in strongly coupled\\ Bogolyubov's oscillator model}
\date{}
\maketitle
\begin{abstract}
We examine classical Bogolyubov's model of a particle coupled to a
heat bath which consists of infinitely many stochastic
oscillators. Bogolyubov's result~\cite{bogolubov} suggests that,
in the stochastic limit, the model exhibits convergence to
thermodynamical equilibrium. It has recently been shown that the
system does attain the equilibrium if the coupling constant is
small enough~\cite{alekseev}. We show that in the case of the
large coupling constant the distribution function
$\rho_{S}(q,p,t)\to 0$ pointwise as $t\to\infty$. This implies
that if there is convergence to equilibrium, then the limit
measure has no finite momenta. Besides, the probability to find
the particle in any finite domain of phase space tends to zero.
This is also true for domains in the coordinate space and in the
momentum space.
\end{abstract}

\section{Introduction}
If two bodies with different temperatures are in contact, they
will eventually have the same temperature. The inverse process of
"temperature separation" does not occur if we do not act on the
system by anything. This phenomenon is referred to as
irreversibility. It seems paradoxical since equations of mechanics
(Newton's equation) and quantum mechanics (Schroedinger's
equation) are time-reversible. This problem has been discussed for
a long time and a lot of outstanding scientists such as Boltzmann,
Poincare, Gibbs, Birkhoff, Bogolyubov and others tried to solve
it. As a result, new approaches and techniques have been developed
\cite{bogolubov}-\cite{Kozlov}. One of the recently developed
techniques is a stochastic limit (see \cite{acc-lu-vol} and
references therein).

 The idea of Bogolyubov's model \cite{bogolubov}, that considers the behavior
of one particular oscillator under the action of many other
stochastic oscillators, was later further
developed~\cite{feynman}. The quantum analogue of Bogolyubov's
model has been studied in details as well
(see~\cite{FLO}-\cite{Ford-1} and references therein).

In this paper we first briefly describe Bogolyubov's model.
Bogolyubov~\cite{bogolubov} suggested a toy model that could
represent a system in contact with a thermostat. The thermostat is
modelled by an infinite number of oscillators whose initial
coordinates and momenta are random variables with thermal (Gibbs)
distribution. The system is represented by a single oscillator whose
coordinate and momentum are arbitrarily fixed at the initial
instant. The system interacts with the thermostat with some coupling
constant. It is expected that asymptotically the system gets the
same temperature as the thermostat, i.e. the coordinate and momentum
of the single oscillator will obey the Gibbs distribution. In his
paper~\cite{bogolubov} Bogolyubov proved an estimate of the
distribution function~$\rho_{S}(q,p,t)$ in some interval of $t$,
which suggests that, in the stochastic limit~\cite{acc-lu-vol}, the
model exhibits convergence to thermodynamical equilibrium.
Bogolyubov's model is simple enough to prove theorems or make
explicit calculations in some particular cases. It has recently been
shown that the system does attain the equilibrium if the coupling
constant is small enough~\cite{alekseev}.

In this paper we examine the model in the case of the large coupling
constant. We find that in this case the distribution function
$\rho_{S}(q,p,t)\to 0$ pointwise as $t\to\infty$. This implies that
if there is convergence to equilibrium, then the limit measure has
no finite momenta and is not the Gibbs function. Besides, the
probability to find the particle in any finite domain of phase space
tends to zero. This is also true for domains in the coordinate space
and in the momentum space.

The outline of the paper is as follows. In Sec. 2 we formulate a
mathematical model and set out Bogolyubov's results. In Sec. 3 we
give a theorem about attaining equilibrium in a particular case of
small coupling constants~\cite{alekseev}. And in Sec. 4 we
consider another particular case of large coupling constants. In
Sec. 5 we discuss the results.

\section{Model and Bogolyubov's results}
\textbf{The Hamiltonian and Hamilton equations.} The following
model is considered. There is an oscillator (the system) and a set
of $N$ oscillators (the thermostat) with the following total
Hamiltonian:
\begin{equation}
\label{hamiltonian}
H=\frac{1}{2}(p^{2}+\omega^{2}q^{2})+\frac{1}{2}\sum_{n=1}^{N}{(p_{n}^{2}+\omega_{n}^{2}q_{n}^{2})}+\varepsilon\sum_{n=1}^{N}\alpha_{n}q_{n}q,
\end{equation}
where $p,~q,~\omega$ and $p_{n},~q_{n},~\omega_{n}$ are momenta,
coordinates and frequencies of the first oscillator and those of
the set of oscillators, respectively; $\varepsilon$ and
$\alpha_{n}$ are positive numbers and play a role of coupling
constants. In what follows we imply $\varepsilon$ talking about a
small or large coupling constant.

The corresponding Hamilton equations are
\begin{equation}
\label{hamilton-eq}
\begin{array}{cccc}
\displaystyle\frac{d^{2}q_{n}}{dt^{2}}+\omega_{n}^{2}q_{n}=-\varepsilon\alpha_{n}q,
& \displaystyle p_{n}=\frac{dq_{n}}{dt}, & p_{n}(0)=P_{n}, & q_{n}(0)=Q_{n},\\
\displaystyle\frac{d^{2}q}{dt^{2}}+\omega^{2}q=-\varepsilon\sum_{n=1}^{N}\alpha_{n}q_{n},
& \displaystyle p=\frac{dq}{dt}, & p(0)=p_{0}, & q(0)=q_{0}.
\end{array}
\end{equation}

The model parameters $\alpha_{n}$, $\omega_{n}$, $P_{n}$, $Q_{n}$,
$p_{0}$, $q_{0}$ satisfy the following conditions. The initial
momentum and coordinate of the system $p_{0},~q_{0}$ are arbitrary
real numbers: $p_{0},~q_{0}\in\mathbb{R}$.

The parameters $\alpha_{n}$ and the frequencies $\omega_{n}$
satisfy the conditions corresponding to transition to a continuous
spectrum as $N\to\infty$:
\begin{equation}
\begin{array}{cc}
\displaystyle\sum_{0<\omega_{n}<\nu}{\frac{\alpha_{n}^{2}}{\omega_{n}^{2}}}\rightarrow\int_{0}^{\nu}J(\tau)d\tau,
&
\displaystyle\sum_{\nu<\omega_{n}}{\frac{\alpha_{n}^{2}}{\omega_{n}^{2}}}\rightarrow\int_{\nu}^{\infty}J(\tau)d\tau
\end{array}
\end{equation}
for $\forall$~$\nu>0$. $J(\nu)$ is a continuous positive function
and $\displaystyle\int_0^{\infty}{J(\nu)d\nu}<\infty$.

The initial momenta and coordinates of the set of oscillators (the
thermostat) $P_{n}$ and $Q_{n}$ are random variables with the
distribution function
\begin{equation}
\rho(\zeta_{n},\theta_{n})=\exp{\left(\frac{\Psi}{kT}-\frac{1}{2kT}\sum_{n=1}^{N}{(\zeta_{n}^{2}+\omega_{n}^{2}\theta_{n}^{2})}\right)}
\end{equation}
such that
$$
\int_{\mathbb{R}^{2n}}{\rho(\zeta_{n},\theta_{n})}d\zeta_{1}\ldots
d\zeta_{N}d\theta_{1}\ldots d\theta_{N}=1,
$$
where $\Psi\in\mathbb{R}$ and $k,~T$ are positive numbers.
Physically, $~k$~and~$T$ are Boltzmann constant and temperature, respectively.\\
\\
\textbf{Bogolubov's results.} Let us introduce new
variables~$E_{n}$~and~$\varphi_{n}$ as follows:
\begin{equation}
\begin{array}{cc}
Q_{n}=\displaystyle\frac{\sqrt{2E_{n}}}{\omega_{n}}\cos{\varphi_{n}},
& P_{n}=-\sqrt{2E_{n}}\sin{\varphi_{n}},
\end{array}
\end{equation}
so that
$E_{n}=\displaystyle\frac{1}{2}(P_{n}^{2}+\omega_{n}^{2}Q_{n}^{2})$
are initial energies. Further, let
\begin{equation}
K_{N}(t)=\sum_{n=1}^{N}\alpha_{n}^{2}\frac{\sin\omega_{n}t}{\omega_{n}},
\end{equation}
\begin{equation}
f_{N}(t)=-\sum_{n=1}^{N}\alpha_{n}\frac{\sqrt{2E_{n}}}{\omega_{n}}\cos{(\omega_{n}t+\varphi_{n})}
\end{equation}
and $v_{N}(t)$ be a solution of the integro-differential equation
\begin{equation}
\left\{
\begin{array}{l}
\displaystyle v_{N}''(t)+\omega^{2}v_{N}(t)=\varepsilon^{2}\int_{0}^{t}K_{N}(t-\tau)v(\tau)d\tau,\\
v_{N}(0)=0,~~~v_{N}'(0)=1.
\end{array}
\right. \label{integr_diff_eq_N}
\end{equation}
Then the solution $q(t),p(t)$ of equations~(\ref{hamilton-eq})
reads~\cite{bogolubov}
\begin{equation}
\label{hamilton-sol}
\begin{array}{c}
q(t)=\displaystyle q_{0}v_{N}'(t)+p_{0}v_{N}(t)+\varepsilon\int_{0}^{t}v_{N}(t-\tau)f_{N}(\tau)d\tau, \\
p(t)=\displaystyle
q_{0}v_{N}''(t)+p_{0}v_{N}'(t)+\varepsilon\int_{0}^{t}v_{N}'(t-\tau)f_{N}(\tau)d\tau.
\end{array}
\end{equation}
Note that the dependance of the solutions $q(t)$ and $p(t)$ on $N$
is implied.

Bogolyubov~\cite{bogolubov} showed that as $N\to\infty$ the
solution $v_{N}(t)$ along with its first and second derivatives
converges uniformly in any finite interval to $v(t)$. The latter
is a solution of the following integro-differential equation:
\begin{equation}
\left\{
\begin{array}{l}
\displaystyle v''(t)+\omega^{2}v(t)=\varepsilon^{2}\int_{0}^{t}Q(t-\tau)v'(\tau)d\tau,\\
v(0)=0,~~~v'(0)=1,
\end{array}
\right. \label{integr_diff_eq}
\end{equation}
where
$$
Q(t)=\int_{0}^{\infty}J(\nu)(1-\cos{\nu t})d\nu.
$$

According to Bogolubov~\cite{bogolubov} we can formulate

\textbf{Theorem 1.} There exists a limit of the probability
density of random values $q(t)$, $p(t)$ for any $t>0$ as
$N\to\infty$:
$$
\rho_{S}(t,q,p)=\Phi(q-q^{*}(t),p-p^{*}(t),t).
$$
The limit is meant in the following sense:
\begin{equation}
\lim_{N\to\infty} Prob\{a_{1}<q(t)<a_{2},b_{1}<p(t)<b_{2}\}=
\int_{a1}^{a2}\int_{b1}^{b2}\rho_{S}(t,\xi,\eta)d\xi d\eta.
\end{equation}
Here
\begin{equation}
q^{*}(t)=q_{0}v'(t)+p_{0}v(t),~~~p^{*}(t)=q_{0}v''(t)+p_{0}v'(t)
\end{equation}
and
\begin{equation}
\Phi(\xi,\eta,t)=\frac{1}{2\pi\sqrt{AC-B^{2}}}\exp{\left(-\frac{C\xi^{2}-2B\xi\eta+A\eta^{2}}{2(AC-B^{2})}\right)}.
\label{Phi_func}
\end{equation}
The coefficients $A=A(t)$, $B=B(t)$ and $C=C(t)$ are derived from
the identity
\begin{equation}
\label{ABC_Def}
A(t)\lambda^{2}+2B(t)\lambda\mu+C(t)\mu^{2}\equiv\varepsilon^{2}kT\int_{0}^{\infty}{J(\nu)}\left|\int_{0}^{t}{\{\lambda
v(x)+\mu v'(x)\}e^{-i\nu x}dx}\right|^{2}d\nu.
\end{equation}
From the identity (\ref{ABC_Def}) it is clear that, first, $A\geq
0$ and, second, $AC-B^{2}>0$. The latter is obvious, because the
right hand side is positive for any $\lambda$ and $\mu$, hence,
the discriminant $4(B^{2}-AC)<0$.

The second important result in~\cite{bogolubov} yields an estimate
of the limit function $\rho_{S}(t,q,p)$ in some interval with
respect to $t$ and is formulated as

\textbf{Theorem 2.} For $\forall\varepsilon>0$,
$\forall\beta>\alpha>0$, and for any sequence $\{\triangle
t_{\varepsilon}\}$ such that $\triangle t_{\varepsilon}\to\infty$,
$\varepsilon^{2}\triangle t_{\varepsilon}\to 0$ as $\varepsilon\to
0$ we have for $\displaystyle\forall
t\in\left(\frac{\alpha}{\varepsilon^{2}},\frac{\beta}{\varepsilon^{2}}\right)$
\begin{equation}
\left|\frac{1}{\triangle t_{\varepsilon}}\int_{t}^{t+\triangle
t_{\varepsilon}}{(\rho_{S}-\rho_{S}^{0})}\right|<\sigma(\varepsilon),
\end{equation}
where $\sigma(\varepsilon)\to 0$ as $\varepsilon\to 0$, and
$\rho_{S}^{0}$ is some explicit expression which tends to the
Gibbs distribution with temperature $T$ as $t\to\infty$.

However, Theorem~$2$ tells us nothing at all about the asymptotic
behavior of $\rho_{S}(t,q,p)$ as $t\to\infty$. In~\cite{alekseev}
a particular case has been considered and some kind of an
asymptotics, which tends to the Gibbs function, has been found.
The respective theorem is formulated in the next section.

\section{Particular case with a small coupling constant $\varepsilon$}
We shall leave clarifying what should be considered as a small or
a large coupling constant till the next section.

\textbf{Theorem 3.} Let $J(\nu)\in C(\mathbb{R})\cap
L_{1}(\mathbb{R})$ be an even rational function, and all its
critical points in $\mathbb{C}$ are of the first order. Then for
any $\sigma>0$ there is $\varepsilon_{0}$ that for any
$\varepsilon$: $0<|\varepsilon|<\varepsilon_{0}$ there exists such
$t_{0}(\varepsilon)$ that when $t>t_{0}(\varepsilon)$ we have for
any $p,q\in\mathbb{R}$
\begin{equation}
\left|\rho_{S}(q,p,t)-\frac{\omega}{2\pi
kT(1-e^{-2\delta(\varepsilon)t})}\cdot
exp\left(-\frac{E+E_{0}e^{-2\delta(\varepsilon)t}-
2\sqrt{EE_{0}}e^{-\delta(\varepsilon)t}\cos{((\omega+\varepsilon^{2}Im\rho)t+
\varphi_{0}-\varphi)}}{(1-e^{-2\delta(\varepsilon)t})kT}\right)\right|<\sigma,
\end{equation}
where $\delta(\varepsilon)$ and $\rho(\varepsilon)$ are determined
by the function $J(\nu)$. Besides,
$$
\begin{array}{cc}
q=\displaystyle\frac{\sqrt{2E}}{\omega}\cos{\varphi}, &
p=-\sqrt{2E}\sin{\varphi}, \\
\\
q_{0}=\displaystyle\frac{\sqrt{2E_{0}}}{\omega}\cos{\varphi_{0}},
&
p_{0}=-\sqrt{2E_{0}}\sin{\varphi_{0}}, \\
\end{array}
$$
i.e. $E=\displaystyle\frac{p^{2}}{2}+\frac{\omega^{2}q^{2}}{2}$ is
the energy.

From Theorem $3$ one can easily see that the asymptotics as
$t\to\infty$ tends to the Gibbs function.

\section{Particular case with a large coupling constant $\varepsilon$}
Let us consider a particular case with
\begin{equation}
\label{J-function}
\begin{array}{lll}
J(\nu)=\displaystyle\frac{1}{a+b\nu^{2}}, & a>0, & b>0.
\end{array}
\end{equation}
Obviously, this function satisfies the Theorem $3$ conditions.

\textbf{Proposition 1.} If $J(\nu)$ is taken as specified
in~(\ref{J-function}) and function $v(t)$ in integro-differential
equation~(\ref{integr_diff_eq}) is triply continuously
differentiable then (\ref{integr_diff_eq}) takes the form of a
third-order differential equation:
\begin{equation}
\label{v-diff-equation}
\left\{
\begin{array}{l}
\upsilon'''(t)+\displaystyle\left(\frac{a}{b}\right)^{1/2}\upsilon''(t)+
\omega^{2}\upsilon'(t)+\left(\left(\frac{a}{b}\right)^{1/2}\omega^{2}-\frac{\varepsilon^{2}\pi}{2b}\right)\upsilon(t)=0,\\
\upsilon(0)=0,~\upsilon'(0)=1,~\upsilon''(0)=0,~0\leq t<+\infty.
\end{array}
\right. \label{diff_eq}
\end{equation}

\textbf{Proof.} First of all, let us calculate $Q(t)$ with
residues.
\begin{equation}
\label{Q-function}
\begin{array}{l}
Q(t)\equiv\displaystyle\int_{0}^{\infty}J(\nu)(1-\cos{\nu t})d\nu=
\int_{0}^{\infty}\frac{1-\cos{\nu t}}{a+b\nu^{2}}d\nu=
\frac{1}{2}\int_{-\infty}^{\infty}\frac{1-e^{i\nu
t}}{a+b\nu^{2}}d\nu= \\
\\
=\pi
i\displaystyle\left[\frac{1}{2bi\sqrt{a/b}}-\frac{e^{-\sqrt{\frac{a}{b}}t}}{2bi\sqrt{a/b}}\right]=
\frac{\pi}{2\sqrt{ab}}\left[1-e^{-\sqrt{\frac{a}{b}}t}\right].
\end{array}
\end{equation}

Integrating the right-hand side of~(\ref{integr_diff_eq}) by parts
we can write it in the following form:
\begin{equation}
v''(t)+\omega^{2}v(t)=\varepsilon^{2}\int_{0}^{t}v(\tau)Q'(t-\tau)d\tau.
\end{equation}
Taking into account the explicit formula~(\ref{Q-function}) we
obtain:
\begin{equation}
\label{v-equation}
v''(t)+\omega^{2}v(t)=\frac{\varepsilon^{2}\pi}{2b}F(t),
\end{equation}
where
$$
F(t)=\int_{0}^{t}v(\tau)\exp{\left(-\sqrt{\frac{a}{b}}
(t-\tau)\right)}d\tau.
$$
Clearly, $F(t)$ satisfies the equation:
\begin{equation}
\frac{dF}{dt}=-\sqrt{\frac{a}{b}}F+v(t).
\end{equation}
Then we differentiate both parts of equation~(\ref{v-equation})
and obtain the third-order differential equation
in~(\ref{v-diff-equation}). One more initial condition, which is
the value of the second derivative $v''(0)$, directly comes
from~(\ref{integr_diff_eq}) if we let $t=0$. The proposition is
proved.

The corresponding characteristic equation
for~(\ref{v-diff-equation}) is
\begin{equation}
\lambda^{3}+\sqrt{\frac{a}{b}}\lambda^{2}+\omega^{2}\lambda+\sqrt{\frac{a}{b}}\omega^{2}-\frac{\varepsilon^{2}\pi}{2b}=0,
\label{charac-equ}
\end{equation}
or
\begin{equation}
\label{charac-equ1}
\left(\lambda^{2}+\omega^{2}\right)\left(\lambda+\sqrt{\frac{a}{b}}\right)=\frac{\varepsilon^{2}\pi}{2b}.
\end{equation}

At this point we can formulate the difference between a small and
a large coupling constant. If equation (\ref{charac-equ}) has two
complex roots, which differ by order of $\varepsilon^{2}$ from
$i\omega$ and $-i\omega$ (the roots are purely imaginary in the
case of $\varepsilon=0$), and one real root which differs by the
same order from $-\sqrt{a/b}$, then this is the case of a small
coupling constant. And this is the case of a large coupling
constant when (\ref{charac-equ}) has three real roots: two
negative and one positive. We can make sure that the
characteristic equation \textit{can} have two negative and one
positive roots. Let
\mbox{$\omega^{2}=1/3$},\mbox{$\varepsilon\pi/2b=4$}~and~$a/b=9$.
Then the characteristic equation~(\ref{charac-equ}) takes the
form:
$$
\left(\lambda^{2}+\frac{1}{3}\right)(\lambda+3)=4.
$$
It is easy to check that the last equation has three real roots
whose approximate values are $-\lambda_{1}\approx -2.2723$,
$-\lambda_{2}\approx -1.5691$ and $\lambda_{3}\approx 0.8414$. In
the case of a large coupling constant we shall prove the following

\textbf{Theorem 4.} Let equation (\ref{charac-equ}) have three
real roots two of which are negative and one is positive:
$-\lambda_{1}$, $-\lambda_{2}$ and $\lambda_{3}$, where
$\lambda_{1}>0$, $\lambda_{2}>0$, $\lambda_{3}>0$. Further, let
$\lambda_{3}<\lambda_{2}$, $\lambda_{3}<\lambda_{1}$ and
$\lambda_{3}<\sqrt{a/b}$. Then the pointwise limit with respect to
$q$ and $p$ is equal to zero:
\begin{equation}
\lim_{t\to+\infty}{\rho_{S}(t,q,p)}=0.
\end{equation}

\textbf{Proof.} In order to prove the theorem we explicitly
calculate $A$, $B$ and $C$. According to the conditions of the
theorem the solution of equation~(\ref{diff_eq}) is
\begin{equation}
v(t)=C_{1}e^{-\lambda_{1}t}+C_{2}e^{-\lambda_{2}t}+C_{3}e^{\lambda_{3}t}.
\end{equation}

From the initial conditions we have:
\begin{equation}
\label{constants}
\begin{array}{c}
C_{1}=\displaystyle\frac{\lambda_{3}-\lambda_{2}}{(\lambda_{2}-\lambda_{1})(\lambda_{1}+\lambda_{3})},\\
~\\
C_{2}=\displaystyle\frac{\lambda_{1}-\lambda_{3}}{(\lambda_{2}-\lambda_{1})(\lambda_{2}+\lambda_{3})},\\
~\\
C_{3}=\displaystyle\frac{\lambda_{1}+\lambda_{2}}{(\lambda_{2}+\lambda_{3})(\lambda_{1}+\lambda_{3})}.
\end{array}
\end{equation}

Then we find~$A(t)$,~$B(t)$ and $C(t)$ from the equality:
\begin{equation}
\label{ABC_Def}
A(t)\lambda^{2}+2B(t)\lambda\mu+C(t)\mu^{2}\equiv\varepsilon^{2}kT\int_{0}^{\infty}{J(\nu)}\left|\int_{0}^{t}{\{\lambda
v(x)+\mu v'(x)\}e^{-i\nu x}dx}\right|^{2}d\nu,
\end{equation}
where~$J(\nu)=\displaystyle\frac{1}{a+b\nu^{2}}$.

Let us introduce $I_{i}$ and $S_{i}$, $i=\overline{1,6}$, as
follows:
\begin{equation}
\left|\int_{0}^{t}{\{\lambda v(x)+\mu v'(x)\}e^{-i\nu
x}dx}\right|^{2}=I_{1}+I_{2}+I_{3}+I_{4}+I_{5}+I_{6},
\end{equation}

\begin{equation}
\int_{0}^{\infty}{J(\nu)}\left|\int_{0}^{t}{\{\lambda v(x)+\mu
v'(x)\}e^{-i\nu
x}dx}\right|^{2}d\nu=S_{1}+S_{2}+S_{3}+S_{4}+S_{5}+S_{6},
\end{equation}
where
$$
S_{i}\equiv\int_{0}^{\infty}{J(\nu)I_{i}d\nu}.
$$

Straightforward, but tedious calculations give (some intermediate
calculations are carried out in Appendix A)
\begin{equation}
I_{1}=C_{1}^{2}\frac{(\lambda-\mu\lambda_{1})^{2}}{\lambda_{1}^{2}+\nu^{2}}\left(1-2e^{-\lambda_{1}t}\cos{\nu
t}+e^{-2\lambda_{1}t}\right).
\end{equation}
\begin{equation}
I_{2}=C_{2}^{2}\frac{(\lambda-\mu\lambda_{2})^{2}}{\lambda_{2}^{2}+\nu^{2}}\left(1-2e^{-\lambda_{2}t}\cos{\nu
t}+e^{-2\lambda_{2}t}\right).
\end{equation}
\begin{equation}
I_{3}=C_{3}^{2}\frac{(\lambda+\mu\lambda_{3})^{2}}{\lambda_{3}^{2}+\nu^{2}}\left(1-2e^{\lambda_{3}t}\cos{\nu
t}+e^{2\lambda_{3}t}\right).
\end{equation}
\begin{equation}
\begin{array}{c}
I_{4}=\displaystyle
2C_{1}C_{2}\frac{(\lambda-\mu\lambda_{1})(\lambda-\mu\lambda_{2})}{(\lambda_{1}^{2}+\nu^{2})(\lambda_{2}^{2}+\nu^{2})}\left[(1+e^{-(\lambda_{1}+\lambda_{2})t})(\lambda_{1}\lambda_{2}+\nu^{2})-\right.\\
~\\
\left.-(e^{-\lambda_{1}t}+e^{-\lambda_{2}t})(\lambda_{1}\lambda_{2}+\nu^{2})\cos{\nu
t
}-\nu(e^{-\lambda_{1}t}-e^{-\lambda_{2}t})(\lambda_{1}-\lambda_{2})\sin{\nu
t }\right].
\end{array}
\end{equation}

\begin{equation}
\begin{array}{c}
I_{5}=\displaystyle
2C_{2}C_{3}\frac{(\lambda+\mu\lambda_{3})(\lambda-\mu\lambda_{2})}{(\lambda_{3}^{2}+\nu^{2})(\lambda_{2}^{2}+\nu^{2})}\left[(1+e^{(\lambda_{3}-\lambda_{2})t})(\nu^{2}-\lambda_{3}\lambda_{2})-\right.\\
~\\
\left.-(e^{\lambda_{3}t}+e^{-\lambda_{2}t})(\nu^{2}-\lambda_{3}\lambda_{2})\cos{\nu
t
}+\nu(e^{\lambda_{3}t}-e^{-\lambda_{2}t})(\lambda_{2}+\lambda_{3})\sin{\nu
t }\right].
\end{array}
\end{equation}

\begin{equation}
\begin{array}{c}
I_{6}=\displaystyle
2C_{1}C_{3}\frac{(\lambda-\mu\lambda_{1})(\lambda+\mu\lambda_{3})}{(\lambda_{1}^{2}+\nu^{2})(\lambda_{3}^{2}+\nu^{2})}\left[(1+e^{(\lambda_{3}-\lambda_{1})t})(\nu^{2}-\lambda_{1}\lambda_{3})-\right.\\
~\\
\left.-(e^{-\lambda_{1}t}+e^{\lambda_{3}t})(\nu^{2}-\lambda_{1}\lambda_{3})\cos{\nu
t
}-\nu(e^{-\lambda_{1}t}-e^{\lambda_{3}t})(\lambda_{3}+\lambda_{1})\sin{\nu
t }\right].
\end{array}
\end{equation}

\begin{equation}
S_{1}=\frac{\pi
C_{1}^{2}(\lambda-\mu\lambda_{1})^{2}}{2(a-b\lambda_{1}^{2})}\left[(1+e^{-2\lambda_{1}
t})\left(\frac{1}{\lambda_{1}}-\sqrt{\frac{b}{a}}\right)-2e^{-\lambda_{1}t}\left(\frac{e^{-\lambda_{1}
t}}{\lambda_{1}}-\sqrt{\frac{b}{a}}e^{-\sqrt{\frac{a}{b}}
t}\right)\right].
\end{equation}

\begin{equation}
S_{2}=\frac{\pi
C_{2}^{2}(\lambda-\mu\lambda_{2})^{2}}{2(a-b\lambda_{2}^{2})}\left[(1+e^{-2\lambda_{2}
t})\left(\frac{1}{\lambda_{2}}-\sqrt{\frac{b}{a}}\right)-2e^{-\lambda_{2}t}\left(\frac{e^{-\lambda_{2}
t}}{\lambda_{2}}-\sqrt{\frac{b}{a}}e^{-\sqrt{\frac{a}{b}}
t}\right)\right].
\end{equation}

\begin{equation}
S_{3}=\frac{\pi
C_{3}^{2}(\lambda+\mu\lambda_{3})^{2}}{2(a-b\lambda_{3}^{2})}\left[(1+e^{2\lambda_{3}
t})\left(\frac{1}{\lambda_{3}}-\sqrt{\frac{b}{a}}\right)-2e^{\lambda_{3}t}\left(\frac{e^{-\lambda_{3}
t}}{\lambda_{3}}-\sqrt{\frac{b}{a}}e^{-\sqrt{\frac{a}{b}}
t}\right)\right].
\end{equation}

\begin{equation}
\begin{array}{c}
S_{4}=\displaystyle\frac{\pi
C_{1}C_{2}(\lambda-\mu\lambda_{1})(\lambda-\mu\lambda_{2})
}{(a-b\lambda_{1}^{2})(a-b\lambda_{2}^{2})}\left[\left(1-e^{-(\lambda_{1}+\lambda_{2})t}\right)\frac{2a-b(\lambda_{1}^{2}+\lambda_{2}^{2})}{\lambda_{1}+\lambda_{2}}+\right.\\
~\\
+\displaystyle\left(\lambda_{1}\lambda_{2}\frac{b^{3/2}}{\sqrt{a}}-\sqrt{ab}\right)\left(1+e^{-(\lambda_{1}+\lambda_{2})t}-e^{-(\lambda_{1}+\sqrt{\frac{a}{b}})t}-
e^{-(\lambda_{2}+\sqrt{\frac{a}{b}})t}\right)-\\
~\\
-\left.
be^{-\sqrt{\frac{a}{b}}t}\left(e^{-\lambda_{2}t}-e^{-\lambda_{1}t}\right)(\lambda_{2}-\lambda_{1})\right].
\end{array}
\end{equation}

\begin{equation}
\begin{array}{c}
S_{5}=\displaystyle\frac{\pi
C_{2}C_{3}(\lambda-\mu\lambda_{2})(\lambda+\mu\lambda_{3})
}{(a-b\lambda_{2}^{2})(a-b\lambda_{3}^{2})}\left[\left(1-e^{(\lambda_{3}-\lambda_{2})t}\right)\frac{2a-b(\lambda_{2}^{2}+\lambda_{3}^{2})}{\lambda_{2}-\lambda_{3}}-\right.\\
~\\
-\displaystyle\left(\lambda_{2}\lambda_{3}\frac{b^{3/2}}{\sqrt{a}}+\sqrt{ab}\right)\left(1+e^{(\lambda_{3}-\lambda_{2})t}-e^{-(\lambda_{2}+\sqrt{\frac{a}{b}})t}-
e^{(\lambda_{3}-\sqrt{\frac{a}{b}})t}\right)-\\
~\\
-\left.
be^{-\sqrt{\frac{a}{b}}t}\left(e^{-\lambda_{2}t}-e^{\lambda_{3}t}\right)(\lambda_{2}+\lambda_{3})\right].
\end{array}
\end{equation}

\begin{equation}
\begin{array}{c}
S_{6}=\displaystyle\frac{\pi
C_{1}C_{3}(\lambda-\mu\lambda_{1})(\lambda+\mu\lambda_{3})
}{(a-b\lambda_{1}^{2})(a-b\lambda_{3}^{2})}\left[\left(1-e^{(\lambda_{3}-\lambda_{1})t}\right)\frac{2a-b(\lambda_{1}^{2}+\lambda_{3}^{2})}{\lambda_{1}-\lambda_{3}}-\right.\\
~\\
-\displaystyle\left(\lambda_{1}\lambda_{3}\frac{b^{3/2}}{\sqrt{a}}+\sqrt{ab}\right)\left(1+e^{(\lambda_{3}-\lambda_{1})t}-e^{-(\lambda_{1}+\sqrt{\frac{a}{b}})t}-
e^{(\lambda_{3}-\sqrt{\frac{a}{b}})t}\right)-\\
~\\
-\left.
be^{-\sqrt{\frac{a}{b}}t}\left(e^{-\lambda_{1}t}-e^{\lambda_{3}t}\right)(\lambda_{1}+\lambda_{3})\right].
\end{array}
\end{equation}

We define $P_{i}$, $i=\overline{1,6}$ as follows:
\begin{equation}
\begin{array}{c}
S_{1}=(\lambda-\mu\lambda_{1})^{2}P_{1},\\
S_{2}=(\lambda-\mu\lambda_{2})^{2}P_{2},\\
S_{3}=(\lambda+\mu\lambda_{3})^{2}P_{3},\\
S_{4}=(\lambda-\mu\lambda_{1})(\lambda-\mu\lambda_{2})P_{4},\\
S_{5}=(\lambda-\mu\lambda_{2})(\lambda+\mu\lambda_{3})P_{5},\\
S_{6}=(\lambda-\mu\lambda_{1})(\lambda+\mu\lambda_{3})P_{6}.\\
\end{array}
\end{equation}

Now we use the theorem conditions $\lambda_{3}<\lambda_{2},
\lambda_{3}<\lambda_{1}, \lambda_{3}<\sqrt{a/b}$.

Then all $S_{i}$ but $S_{3}$ tend to constants. The latter grows
exponentially: $S_{3}\propto e^{2\lambda_{3}t}$. (Hereafter, the
notation $\propto$ has the same meaning as in $S_{3}=const\cdot
e^{2\lambda_{3}t}+o(e^{2\lambda_{3}t})$ as $t\to\infty$).

From definition (\ref{ABC_Def}) we have:
\begin{equation}
\label{ABC-view}
\begin{array}{l}
\displaystyle\frac{A}{\varepsilon^{2}kT}=P_{1}+P_{2}+P_{3}+P_{4}+P_{5}+P_{6},\\~\\
\displaystyle
\displaystyle\frac{B}{\varepsilon^{2}kT}=-\lambda_{1}P_{1}-\lambda_{2}P_{2}+\lambda_{3}P_{3}-\frac{1}{2}(\lambda_{1}+\lambda_{2})P_{4}-\frac{1}{2}(\lambda_{2}-\lambda_{3})P_{5}-\frac{1}{2}(\lambda_{1}-\lambda_{3})P_{6},\\~\\
\displaystyle\frac{C}{\varepsilon^{2}kT}=\lambda_{1}^{2}P_{1}+\lambda_{2}^{2}P_{2}+\lambda_{3}^{2}P_{3}+\lambda_{1}\lambda_{2}P_{4}-\lambda_{2}\lambda_{3}P_{5}-\lambda_{1}\lambda_{3}P_{6},
\end{array}
\label{coef_behave}
\end{equation}
where $P_{3}\propto e^{2\lambda_{3}t}$ and the other $P_{i}\propto
const$ as $t\rightarrow +\infty$.

Then, we investigate the behavior of $AC-B^{2}$. It is clear that
the terms quadratic in $P_{3}$ are eliminated.  The question is
whether the coefficient in front of terms linear in $P_{3}$ is
zero or not.

Making necessary substitutions from~(\ref{ABC-view}) we obtain:
\begin{equation}
\begin{array}{c}
\displaystyle
\displaystyle\frac{AC-B^{2}}{(\varepsilon^{2}kT)^{2}}=-\frac{1}{4}(\lambda_{1}-\lambda_{2})^{2}P_{4}^{2}-\frac{1}{4}(\lambda_{2}+\lambda_{3})^{2}P_{5}^{2}-\frac{1}{4}(\lambda_{1}+\lambda_{3})^{2}P_{6}^{2}+\\
~\\
+(\lambda_{1}-\lambda_{2})^{2}P_{1}P_{2}+(\lambda_{1}+\lambda_{3})^{2}P_{1}P_{3}+(\lambda_{1}+\lambda_{3})(\lambda_{1}-\lambda_{2})P_{1}P_{5}+\\~\\
+(\lambda_{2}+\lambda_{3})^{2}P_{2}P_{3}+(\lambda_{2}+\lambda_{3})(\lambda_{2}-\lambda_{1})P_{2}P_{6}+(\lambda_{2}+\lambda_{3})(\lambda_{1}+\lambda_{3})P_{3}P_{4}+\\~
\\
\displaystyle
+\frac{1}{2}(\lambda_{1}-\lambda_{2})(\lambda_{2}+\lambda_{3})P_{4}P_{5}+\frac{1}{2}(\lambda_{2}-\lambda_{1})(\lambda_{1}+\lambda_{3})P_{4}P_{6}-
\frac{1}{2}(\lambda_{1}+\lambda_{3})(\lambda_{2}+\lambda_{3})P_{5}P_{6}.
\end{array}
\end{equation}

From the latter expression we obtain that the behavior is as
follows as $t\rightarrow+\infty$:
\begin{equation}
\begin{array}{c}
AC-B^{2}\propto\alpha P_{3},\\
\alpha=(\lambda_{1}+\lambda_{3})^{2}P_{1}+(\lambda_{2}+\lambda_{3})^{2}P_{2}+(\lambda_{2}+\lambda_{3})(\lambda_{1}+\lambda_{3})P_{4}.
\end{array}
\label{P3_behave}
\end{equation}

\textbf{Proposition 2.}
$\displaystyle\lim_{t\to\infty}{\alpha>0}$.

\textbf{Proof.} The behaviors of $P_{1}$, $P_{2}$, $P_{4}$ are:
\begin{equation}
\begin{array}{c}
\displaystyle P_{1}\propto\frac{\pi
C_{1}^{2}}{2(a-b\lambda_{1}^{2})}\left(\frac{1}{\lambda_{1}}-\sqrt{\frac{b}{a}}\right),\\~\\
\displaystyle P_{2}\propto\frac{\pi
C_{2}^{2}}{2(a-b\lambda_{2}^{2})}\left(\frac{1}{\lambda_{2}}-\sqrt{\frac{b}{a}}\right),\\~\\
\displaystyle P_{4}\propto\frac{\pi
C_{1}C_{2}}{(a-b\lambda_{1}^{2})(a-b\lambda_{2}^{2})}\left(\frac{2a-b(\lambda_{1}^{2}+\lambda_{2}^{2})}{\lambda_{1}+\lambda_{2}}+
\lambda_{1}\lambda_{2}\frac{b^{3/2}}{\sqrt{a}}-\sqrt{ab}\right).
\end{array}
\end{equation}

The Viet theorem for characteristic equation~(\ref{charac-equ})
takes the form:
\begin{equation}
\label{Viet} \left\{
\begin{array}{l}
\displaystyle\lambda_{3}-\lambda_{1}-\lambda_{2}=-\sqrt{\frac{a}{b}},\\
\lambda_{1}\lambda_{2}-\lambda_{2}\lambda_{3}-\lambda_{1}\lambda_{3}=\omega^{2},\\
\displaystyle\lambda_{1}\lambda_{2}\lambda_{3}=\frac{\varepsilon^{2}\pi}{2b}-\sqrt{\frac{a}{b}}\omega^{2}.
\end{array}
\right.
\end{equation}

An intermediate result is
$\displaystyle\lim_{t\to\infty}\alpha=\beta R$ (the Viet
theorem~(\ref{Viet}) is already partially used and
relations~(\ref{constants}) are taken into account), where
\begin{equation}
\begin{array}{c}
\displaystyle R=a\left[\frac{(\lambda_{3}-\lambda_{2})^{2}}{\lambda_{1}}+\frac{(\lambda_{3}-\lambda_{1})^{2}}{\lambda_{2}}-\frac{4(\lambda_{3}-\lambda_{2})(\lambda_{3}-\lambda_{1})}{\lambda_{1}+\lambda_{2}}\right]-\\~\\
\displaystyle
-b\left[\frac{\lambda_{2}^{2}}{\lambda_{1}}(\lambda_{3}-\lambda_{2})^{2}+\frac{\lambda_{1}^{2}}{\lambda_{2}}(\lambda_{3}-\lambda_{1})^{2}-2\frac{(\lambda_{1}^{2}+\lambda_{2}^{2})(\lambda_{3}-\lambda_{2})(\lambda_{3}-\lambda_{1})}{\lambda_{1}+\lambda_{2}}\right].
\end{array}
\end{equation}
\begin{equation}
\beta=\frac{\pi}{2(a-b\lambda_{1}^{2})(a-b\lambda_{2}^{2})(\lambda_{2}-\lambda_{1})^{2}}.
\end{equation}

Using Viet relations (\ref{Viet}) again we have:
\begin{equation}
\begin{array}{c}
R=a\displaystyle\frac{(\lambda_{2}-\lambda_{1})^{2}}{\lambda_{1}\lambda_{2}(\lambda_{1}+\lambda_{2})}\left(\lambda_{1}\lambda_{2}+\frac{a}{b}\right)+\\~\\
\displaystyle-b\frac{\lambda_{1}^{2}+\lambda_{2}^{2}}{2}\frac{(\lambda_{2}-\lambda_{1})^{2}}{\lambda_{1}\lambda_{2}(\lambda_{1}+\lambda_{2})}\left(\lambda_{1}\lambda_{2}+\frac{a}{b}\right)+
b\frac{(\lambda_{2}-\lambda_{1})^{2}(\lambda_{2}+\lambda_{1})^{2}}{2\lambda_{1}\lambda_{2}(\lambda_{1}+\lambda_{2})}\left(\lambda_{1}\lambda_{2}-\frac{a}{b}\right).
\end{array}
\end{equation}

Hence,
\begin{equation}
\label{final}
\displaystyle\lim_{t\to\infty}\alpha=\frac{\pi}{b\lambda_{1}\lambda_{2}(\lambda_{1}+\lambda_{2})}>0,
\end{equation}
which proves the proposition.

\textbf{Calculating the exponent in (\ref{Phi_func}).} The
exponent we calculate is:
\begin{equation}
\Pi=-\frac{C\xi^{2}-2B\xi\eta+A\eta^{2}}{2(AC-B^{2})},
\end{equation}
where
$$
\xi=q-q^{*}(t),~~~\eta=p-p^{*}(t)
$$
and
$$
q^{*}(t)=q_{0}v'(t)+p_{0}v(t),~~~p^{*}(t)=q_{0}v''(t)+p_{0}v'(t).
$$

As $t\to\infty$, $q^{*}(t)$ and $p^{*}(t)$ behave as follows:
\begin{equation}
\begin{array}{c}
q^{*}(t)\propto C_{3}e^{\lambda_{3}t}(q_{0}\lambda_{3}+p_{0})\\
p^{*}(t)\propto
\lambda_{3}C_{3}e^{\lambda_{3}t}(q_{0}\lambda_{3}+p_{0}).
\end{array}
\end{equation}
Hence, for arbitrary finite $p$ and $q$
\begin{equation}
\label{xi-eta-lim}
\begin{array}{c}
\xi(t)\propto -C_{3}e^{\lambda_{3}t}(q_{0}\lambda_{3}+p_{0})\\
\eta(t)\propto
-\lambda_{3}C_{3}e^{\lambda_{3}t}(q_{0}\lambda_{3}+p_{0}).
\end{array}
\end{equation}
It yields
\begin{equation}
\Pi=-\frac{C\xi^{2}-2B\xi\eta+A\eta^{2}}{2(AC-B^{2})}\propto
-\frac{(q_{0}\lambda_{3}+p_{0})^{2}C_{3}^{2}e^{2\lambda_{3}t}(C-2B\lambda_{3}+A\lambda_{3}^{2})}{2(AC-B^{2})}.
\end{equation}
Using (\ref{coef_behave}) and(\ref{P3_behave}) we obtain:
\begin{equation}
\lim_{t\to+\infty}{\Pi}=\Pi_{0}=\frac{2a\lambda_{3}^{2}}{\varepsilon^{2}kT\pi}\left(\frac{1}{\lambda_{3}}+\sqrt{\frac{b}{a}}\right)(q_{0}\lambda_{3}+p_{0})^{2},~~~\Pi_{0}=const.
\end{equation}
And finally,
\begin{equation}
\begin{array}{c}
\displaystyle\lim_{t\to+\infty}{\rho_{S}(t,q,p)}|_{p,q=const}=\lim_{t\to+\infty}{\Phi(q-q^{*}(t),p-p^{*}(t),t)}=
\lim_{t\to+\infty}{\frac{1}{2\pi\sqrt{AC-B^{2}}}e^{-\Pi}}=\\
=\displaystyle\lim_{t\to+\infty} { \frac{ e^{-\lambda_{3}t} } {
2\pi\sqrt{\alpha}\varepsilon^{2}kT } \sqrt{ \frac{
2a\lambda_{3}^{2} } { \pi
C_{3}^{2}}\left(\frac{1}{\lambda_{3}}+\sqrt{\frac{b}{a}}\right) }
e^{-\Pi_{0} }}=0,
\end{array}
\end{equation}
which proves the theorem.

\textbf{Corollary 1.} Let
$\rho_{p}(p,t)\equiv\displaystyle\int_{\mathbb{R}}{\rho_{S}(q,p,t)dq}$
and
$\rho_{q}(q,t)\equiv\displaystyle\int_{\mathbb{R}}{\rho_{S}(q,p,t)dp}$.
Then\\
${\displaystyle\lim_{t\to\infty}\rho_{p}(p,t)=\displaystyle\lim_{t\to\infty}\rho_{q}(q,t)=0}$.

\textbf{Proof.} From the explicit expression (\ref{Phi_func}) one
can easily calculate
\begin{equation}
\rho_{p}(p,t)=\frac{1}{\sqrt{2\pi
C}}\exp{\left(-\frac{(p-p^{*}(t))^{2}}{2C}\right)}
\end{equation}
and
\begin{equation}
\rho_{q}(q,t)=\frac{1}{\sqrt{2\pi
A}}\exp{\left(-\frac{(q-q^{*}(t))^{2}}{2A}\right)}.
\end{equation}
Then equations~(\ref{ABC-view})~and~(\ref{xi-eta-lim}) yield:
\begin{equation}
\lim_{t\to\infty}\rho_{p}(p,t)=\lim_{t\to\infty}\frac{1}{\varepsilon\lambda_{3}\sqrt{2\pi
kT}\sqrt{P_{3}}}
\exp{\left(-\frac{\lambda_{3}^{2}C_{3}^{2}(q_{0}\lambda_{3}+p_{0})e^{2\lambda_{3}t}}{2\varepsilon^{2}kT\lambda_{3}^{2}P_{3}}\right)}=0
\end{equation}
since $P_{3}\propto e^{2\lambda_{3}t}$ as $t\to\infty$.
Analogously, $\displaystyle\lim_{t\to\infty}\rho_{q}(q,t)=0$.

\section{Discussion}
Let us calculate the mean coordinate $\langle q\rangle$ and
momentum $\langle p\rangle$, and their standard deviations. Using
{Corollary $1$} we obtain
\begin{equation}
\begin{array}{cc}
\langle
q\rangle=\displaystyle\int_{\mathbb{R}}\theta\rho_{q}(\theta,t)d\theta=q^{*}(t),
& \langle
p\rangle=\displaystyle\int_{\mathbb{R}}\theta\rho_{p}(\theta,t)d\theta=p^{*}(t),\\
\\
\langle
(q-q^{*}(t))^{2}\rangle=\displaystyle\int_{\mathbb{R}}(\theta-q^{*}(t))^{2}\rho_{q}(\theta,t)d\theta=A(t),
& \langle
(p-p^{*}(t))^{2}\rangle=\displaystyle\int_{\mathbb{R}}(\theta-p^{*}(t))^{2}\rho_{p}(\theta,t)d\theta=C(t).
\end{array}
\end{equation}
Again, taking relations~(\ref{ABC-view})~and~(\ref{xi-eta-lim})
into account we see that the behavior of the mean values and the
standard deviations is exponential:
\begin{equation}
\begin{array}{cc}
\langle q\rangle\propto
C_{3}e^{\lambda_{3}t}(q_{0}\lambda_{3}+p_{0}), & \langle
p\rangle\propto
\lambda_{3}C_{3}e^{\lambda_{3}t}(q_{0}\lambda_{3}+p_{0}), \\
\\
\langle (q-q^{*}(t))^{2}\rangle\propto e^{2\lambda_{3}t}, &
\langle (p-p^{*}(t))^{2}\rangle\propto e^{2\lambda_{3}t}.
\end{array}
\end{equation}
This behavior seems strange. Since $q_{0}$ and $p_{0}$ are
arbitrary real numbers, $\langle q\rangle$ and $\langle p\rangle$
may tend either to the positive or negative infinity depending on
$sign{(q_{0}\lambda_{3}+p_{0})}$. It appears that the particle
goes away to the infinity exponentially, and its standard
deviation increases exponentially as well. However, this strange
behavior is explained by

\textbf{Theorem 5.} Characteristic equation~(\ref{charac-equ}) has
a positive root if, and only if, the
Hamiltonian~(\ref{hamiltonian}) is not positive-definite as a
quadratic form of $(2N+1)$ variables
$(q,q_{1},\ldots,q_{N},p_{1},\ldots,p_{N})$ as ${N\to\infty}$.

\textbf{Proof.} We use the Silvester criterion to find out when
the quadratic form $H(q,q_{1},\ldots,q_{N},p_{1},\ldots,p_{N})$
from~(\ref{hamiltonian}) is positive-definite. Its doubled matrix
is
\begin{equation}
\left(
\begin{array}{ccccccccc}
\omega^{2} & \varepsilon\alpha_{1} & \varepsilon\alpha_{1} &
\ldots & \varepsilon\alpha_{N} & 0 &  0 & \ldots & 0 \\
\varepsilon\alpha_{1} & \omega_{1}^{2} & 0 & \ldots & 0 & 0 & 0 &
\ldots & 0 \\
\varepsilon\alpha_{2} & 0 & \omega_{2}^{2} & \ldots & 0 & 0 & 0 &
\ldots & 0 \\
\vdots & \vdots & \vdots & \ddots & 0 & 0 & 0 & \ldots & 0 \\
\varepsilon\alpha_{N} & 0 & 0 & \ldots & \omega_{N}^{2} & 0 & 0 &
\ldots &
0 \\
0 & 0 & 0 & \ldots & 0 & 1 & 0 & \ldots & 0 \\
\vdots & \vdots & \vdots & \ddots & 0 & 0 & 1 & \ldots & 0\\
\vdots & \vdots & \vdots & \vdots & \vdots & 0 & 0 & \ddots & 0\\
0 & 0 & 0 & 0 & 0 & 0 & 0 & \ldots & 1
\end{array}
\right)
\end{equation}
In this case it is sufficient to consider only first $(N+1)$
determinants in the Silvester criterion. The $n$-th determinant
$D_{n}$ can be easily calculated and is as follows:
\begin{equation}
D_{n}=\omega_{1}^{2}\cdot\ldots\cdot\omega_{n-1}^{2}\left(\omega^{2}-
\varepsilon^{2}\sum_{i=1}^{n}{\frac{\alpha^{2}_{i}}{\omega^{2}_{i}}}\right).
\end{equation}
From the latter formula one can see that it is sufficient to
require only $D_{N}$ to be positive: then the rest determinants
are positive. If $D_{N}>0$ then
\begin{equation}
\label{eps-bound}
\varepsilon^{2}<\frac{\omega^{2}}{\displaystyle\sum_{i=1}^{N}{\frac{\alpha^{2}_{i}}{\omega^{2}_{i}}}}.
\end{equation}
Hence, as $N\to\infty$ (\ref{eps-bound}) turns into
\begin{equation}
\label{eps-bound1}
\varepsilon^{2}\leq\frac{\omega^{2}}{\displaystyle\int_{0}^{\infty}J(\tau)d\tau}=\frac{2\sqrt{ba}\omega^{2}}{\pi}.
\end{equation}
Then the right-hand side of~(\ref{charac-equ1}) (which is
equivalent to~(\ref{charac-equ})) is less than
$\displaystyle\omega^{2}\sqrt{\frac{a}{b}}$. Then from the plot of
the left-hand side of~(\ref{charac-equ1}) (as a function of
$\lambda$) it is clear that if (\ref{eps-bound1}) is true,
equation~(\ref{charac-equ}) cannot have a positive root. And vice
versa, if (\ref{eps-bound1}) is not satisfied eq.
(\ref{charac-equ}) has a positive root, but the Hamiltonian is not
positive-definite. Similar divergencies associated with
non-positivity of the density matrix were found in the quantum
analogue of Bogolyubov's model~{\cite{Ford-1}.

It is worth noting that the exponential runaway of the particle
mean coordinate and momentum is not intrinsic to the stochastic
character of the thermal bath oscillators. In the deterministic
case (when one solves~(\ref{hamilton-eq}) with certain initial
data) this also may occur. Indeed, in the simplest case when
$E_{n}=0$ (or, equivalently, $P_{n}=0$ and $Q_{n}=0$) it is easy
to notice from~(\ref{hamilton-sol}) (in this case $f_{N}(t)\equiv
0$) that $\displaystyle\lim_{N\to\infty}q(t)=q^{*}(t)$, and as we
have already seen $q^{*}(t)\propto
C_{3}(\lambda_{3}q_{0}+p_{0})e^{\lambda_{3}t}$.

\section{Conclusion}
It is possible for any coupling constant $\varepsilon$ find such
$a$, $b$ and $\omega$ that the limit (the limit $N\to\infty$ is
computed) distribution function tends to zero as $t\to+\infty$ and
$p$ and $q$ are fixed. This implies that if there is convergence
to equilibrium, then the limit measure has no finite momenta.
Moreover, the probability to find the particle in any finite
domain of the phase, coordinate or momentum space tends to zero,
although the integral all over the space equals to $1$. This
phenomenon might be related to the fact that, as it follows from
{Theorem $5$}, the Hamiltonian is not positive-definite in this
regime for large $N$.

\section{Acknowledgements}
The author is grateful to Prof. I.V.~Volovich for fruitful
discussions and important remarks as well as for raising new
problems associated with Bogolyubov's model. I am also grateful to
\mbox{Prof. L. Accardi} for helpful suggestions. I am thankful to
the referee for useful remarks. The author appreciates discussions
with the participants of the Seminar On The Irreversibility Problem
under the direction of Prof.~V.V.~Kozlov and Prof. I.V.~Volovich.
The work was partially supported by UNK (Study and Science Center)
of the P.N.~Lebedev Physical Institute of Russian Academy of
Sciences.

\appendix

\section{Intermediate calculations of $S_{i}$, $i=4,5,6$}

\begin{equation}
\begin{array}{c}
S_{4}=\displaystyle
C_{1}C_{2}(\lambda-\mu\lambda_{1})(\lambda-\mu\lambda_{2})\left[(1+e^{-(\lambda_{1}+\lambda_{2})t})S_{4,1}-
(e^{-\lambda_{1}t}+e^{-\lambda_{2}t})S_{4,2}-\right.\\
~\\
\left.-(e^{-\lambda_{2}t}-
e^{-\lambda_{1}t})(\lambda_{2}-\lambda_{1})S_{4,3}\right],
\end{array}
\end{equation}
where
\begin{equation}
S_{4,1}=\int_{-\infty}^{\infty}{J(\nu)\frac{\lambda_{1}\lambda_{2}+\nu^{2}}{(\lambda_{1}^{2}+\nu^{2})(\lambda_{2}^{2}+\nu^{2})}d\nu},
\end{equation}

\begin{equation}
S_{4,2}=\int_{-\infty}^{\infty}{J(\nu)\frac{(\lambda_{1}\lambda_{2}+\nu^{2})\cos{\nu
t}}{(\lambda_{1}^{2}+\nu^{2})(\lambda_{2}^{2}+\nu^{2})}d\nu},
\end{equation}

\begin{equation}
S_{4,3}=\int_{-\infty}^{\infty}{J(\nu)\frac{\nu\sin{\nu t
}}{(\lambda_{1}^{2}+\nu^{2})(\lambda_{2}^{2}+\nu^{2})}d\nu}.
\end{equation}

\begin{equation}
S_{4,1}=\frac{\pi}{(a-b\lambda_{1}^{2})(a-b\lambda_{2}^{2})}\left[\frac{2a-b(\lambda_{1}^{2}+\lambda_{2}^{2})}{\lambda_{1}+\lambda_{2}}-\sqrt{ab}+
\lambda_{1}\lambda_{2}\frac{b^{3/2}}{\sqrt{a}}\right].
\end{equation}

\begin{equation}
\begin{array}{c}
S_{4,2}=\displaystyle\frac{\pi}{(a-b\lambda_{1}^{2})(a-b\lambda_{2}^{2})}\left[\frac{a(e^{-\lambda_{1}t}+e^{-\lambda_{2}t})-b(\lambda_{1}^{2}e^{-\lambda_{2}t}+\lambda_{2}^{2}e^{-\lambda_{1}t})}{\lambda_{1}+\lambda_{2}}
+\right.\\
~\\
\left.+
\displaystyle\left(\lambda_{1}\lambda_{2}\frac{b^{3/2}}{\sqrt{a}}-\sqrt{ab}\right)e^{-\sqrt{\frac{a}{b}}t}\right].
\end{array}
\end{equation}

\begin{equation}
S_{4,3}=\frac{\pi}{(a-b\lambda_{1}^{2})(a-b\lambda_{2}^{2})}\left[be^{-\sqrt{\frac{a}{b}}t}+\frac{a(e^{-\lambda_{1}t}-e^{-\lambda_{2}t})}{\lambda_{2}^{2}-\lambda_{1}^{2}}-
\frac{b(\lambda_{2}^{2}e^{-\lambda_{1}t}-\lambda_{1}^{2}e^{-\lambda_{2}t})}{\lambda_{2}^{2}-\lambda_{1}^{2}}\right].
\end{equation}

\begin{equation}
\begin{array}{c}
S_{4}=\displaystyle\frac{\pi
C_{1}C_{2}(\lambda-\mu\lambda_{1})(\lambda-\mu\lambda_{2})
}{(a-b\lambda_{1}^{2})(a-b\lambda_{2}^{2})}\left[\left(1-e^{-(\lambda_{1}+\lambda_{2})t}\right)\frac{2a-b(\lambda_{1}^{2}+\lambda_{2}^{2})}{\lambda_{1}+\lambda_{2}}+\right.\\
~\\
+\displaystyle\left(\lambda_{1}\lambda_{2}\frac{b^{3/2}}{\sqrt{a}}-\sqrt{ab}\right)\left(1+e^{-(\lambda_{1}+\lambda_{2})t}-e^{-(\lambda_{1}+\sqrt{\frac{a}{b}})t}-
e^{-(\lambda_{2}+\sqrt{\frac{a}{b}})t}\right)-\\
~\\
-\left.
be^{-\sqrt{\frac{a}{b}}t}\left(e^{-\lambda_{2}t}-e^{-\lambda_{1}t}\right)(\lambda_{2}-\lambda_{1})\right].
\end{array}
\end{equation}

\begin{equation}
\begin{array}{c}
S_{5}=\displaystyle
C_{2}C_{3}(\lambda-\mu\lambda_{2})(\lambda+\mu\lambda_{3})\left[(1+e^{(\lambda_{3}-\lambda_{2})t})S_{5,1}-
(e^{-\lambda_{2}t}+e^{\lambda_{3}t})S_{5,2}-\right.\\
~\\
\left.-(e^{-\lambda_{2}t}-
e^{\lambda_{3}t})(\lambda_{2}+\lambda_{3})S_{5,3}\right],
\end{array}
\end{equation}
where
\begin{equation}
S_{5,1}=\int_{-\infty}^{\infty}{J(\nu)\frac{\nu^{2}-\lambda_{2}\lambda_{3}}{(\lambda_{2}^{2}+\nu^{2})(\lambda_{3}^{2}+\nu^{2})}d\nu},
\end{equation}

\begin{equation}
S_{5,2}=\int_{-\infty}^{\infty}{J(\nu)\frac{(\nu^{2}-\lambda_{2}\lambda_{3})\cos{\nu
t}}{(\lambda_{2}^{2}+\nu^{2})(\lambda_{3}^{2}+\nu^{2})}d\nu},
\end{equation}

\begin{equation}
S_{5,3}=\int_{-\infty}^{\infty}{J(\nu)\frac{\nu\sin{\nu t
}}{(\lambda_{2}^{2}+\nu^{2})(\lambda_{3}^{2}+\nu^{2})}d\nu}.
\end{equation}

\begin{equation}
S_{5,1}=\frac{\pi}{(a-b\lambda_{2}^{2})(a-b\lambda_{3}^{2})}\left[b(\lambda_{2}+\lambda_{3})-\left(\sqrt{ab}+
\lambda_{2}\lambda_{3}\frac{b^{3/2}}{\sqrt{a}}\right)\right].
\end{equation}

\begin{equation}
\begin{array}{c}
S_{5,2}=\displaystyle\frac{\pi}{(a-b\lambda_{2}^{2})(a-b\lambda_{3}^{2})}\left[a\frac{e^{-\lambda_{3}t}-e^{-\lambda_{2}t}}{\lambda_{3}-\lambda_{2}}+
b\frac{\lambda_{3}^{2}e^{-\lambda_{2}t}-\lambda_{2}^{2}e^{-\lambda_{3}t}}{\lambda_{3}-\lambda_{2}}
-\right.\\
~\\
\left.-
\displaystyle\left(\sqrt{ab}+\lambda_{2}\lambda_{3}\frac{b^{3/2}}{\sqrt{a}}\right)e^{-\sqrt{\frac{a}{b}}t}\right].
\end{array}
\end{equation}

\begin{equation}
S_{5,3}=\frac{\pi}{(a-b\lambda_{2}^{2})(a-b\lambda_{3}^{2})}\left[be^{-\sqrt{\frac{a}{b}}t}+\frac{a(e^{-\lambda_{2}t}-e^{-\lambda_{3}t})}{\lambda_{3}^{2}-\lambda_{2}^{2}}-
\frac{b(\lambda_{3}^{2}e^{-\lambda_{2}t}-\lambda_{2}^{2}e^{-\lambda_{3}t})}{\lambda_{3}^{2}-\lambda_{2}^{2}}\right].
\end{equation}

\begin{equation}
\begin{array}{c}
S_{5}=\displaystyle\frac{\pi
C_{2}C_{3}(\lambda-\mu\lambda_{2})(\lambda+\mu\lambda_{3})
}{(a-b\lambda_{2}^{2})(a-b\lambda_{3}^{2})}\left[\left(1-e^{(\lambda_{3}-\lambda_{2})t}\right)\frac{2a-b(\lambda_{2}^{2}+\lambda_{3}^{2})}{\lambda_{2}-\lambda_{3}}-\right.\\
~\\
-\displaystyle\left(\lambda_{2}\lambda_{3}\frac{b^{3/2}}{\sqrt{a}}+\sqrt{ab}\right)\left(1+e^{(\lambda_{3}-\lambda_{2})t}-e^{-(\lambda_{2}+\sqrt{\frac{a}{b}})t}-
e^{(\lambda_{3}-\sqrt{\frac{a}{b}})t}\right)-\\
~\\
-\left.
be^{-\sqrt{\frac{a}{b}}t}\left(e^{-\lambda_{2}t}-e^{\lambda_{3}t}\right)(\lambda_{2}+\lambda_{3})\right].
\end{array}
\end{equation}

One can see that $S_{6}$ is obtained from $S_{5}$ by
substitution~$2\rightarrow 1$:
\begin{equation}
\begin{array}{c}
S_{6}=\displaystyle\frac{\pi
C_{1}C_{3}(\lambda-\mu\lambda_{1})(\lambda+\mu\lambda_{3})
}{(a-b\lambda_{1}^{2})(a-b\lambda_{3}^{2})}\left[\left(1-e^{(\lambda_{3}-\lambda_{1})t}\right)\frac{2a-b(\lambda_{1}^{2}+\lambda_{3}^{2})}{\lambda_{1}-\lambda_{3}}-\right.\\
~\\
-\displaystyle\left(\lambda_{1}\lambda_{3}\frac{b^{3/2}}{\sqrt{a}}+\sqrt{ab}\right)\left(1+e^{(\lambda_{3}-\lambda_{1})t}-e^{-(\lambda_{1}+\sqrt{\frac{a}{b}})t}-
e^{(\lambda_{3}-\sqrt{\frac{a}{b}})t}\right)-\\
~\\
-\left.
be^{-\sqrt{\frac{a}{b}}t}\left(e^{-\lambda_{1}t}-e^{\lambda_{3}t}\right)(\lambda_{1}+\lambda_{3})\right].
\end{array}
\end{equation}

\end{document}